# Systematic Investigation of Anisotropic Magneto-Peltier Effect and Anomalous Ettingshausen Effect in Ni Thin Films

Raja Das[1,†,‡], Ryo Iguchi[1], and Ken-ichi Uchida[1,2,3,*]

[1]*National Institute for Materials Science, Tsukuba 305-0047, Japan*
[2]*Center for Spintronics Research Network, Tohoku University, Sendai 980-8577, Japan*
[3]*Department of Mechanical Engineering, The University of Tokyo, Tokyo 113-8656, Japan*

[*]UCHIDA.Kenichi@nims.go.jp
[†]Present address: Faculty of Materials Science and Engineering, Phenikaa Institute for Advanced study, Phenikaa University, Hanoi 10000, Vietnam
[‡]Present address: Phenikaa Research and Technology Institute, A&A Green Phoenix Group, Hanoi 10000, Vietnam

The anisotropic magneto-Peltier effect (AMPE) and anomalous Ettingshausen effect (AEE) have been investigated in U-shaped Ni thin films of varying thickness and substrate by means of the lock-in thermography (LIT) method. We have established a procedure to extract pure AMPE and AEE contributions, separated from other thermoelectric effects, for ferromagnetic thin films. The measurements of the magnetic-field-angle $\theta_H$ dependence of the LIT images clearly show that the temperature modulation induced by the AMPE (AEE) in the Ni films varies with the $\cos2\theta_H$ ($\cos\theta_H$) pattern, confirming the symmetry of the AMPE (AEE). The systematic LIT measurements using various substrates show that the AMPE-induced temperature modulation decreases with the increase in thermal conductivity of the substrates, whereas the AEE-induced temperature modulation is almost independent of the thermal conductivity, indicating that the heat loss into the substrates plays an important role in determining the magnitude of the AMPE-induced temperature modulation in thin films. Our experimental results were reproduced by numerical calculations based on a two-dimensional finite element method. These findings provide a platform for investigating the AMPE and AEE in thin film devices.

## I. INTRODUCTION

The Seebeck and Peltier effects are the fundamental thermoelectric phenomena in electric conductors [1,2]. The Seebeck effect, which was discovered by Thomas Johann Seebeck in 1821, refers to the conversion of a heat current into a charge current, enabling thermoelectric voltage generation from waste heat. The Peltier effect, which was discovered by Jean Charles Athanase Peltier in 1834, is the Onsager reciprocal of the Seebeck effect. This phenomenon refers to the conversion of a charge current into a heat current, and generates temperature increase and decrease at the junction of two conductors due to the difference in their Peltier coefficients. Since the Peltier effect enables direct electrical cooling or heating, it is used in solid-state heat pumps and temperature controllers. In these fundamental thermoelectric conversion phenomena, the charge and heat currents flow parallel to each other.

In magnetic materials, in addition to the Seebeck and Peltier effects, a variety of thermoelectric and thermo-spin effects are induced by the concerted action of spin-polarized electron transport and spin-orbit interaction [3-46]. Some of these thermoelectric conversion phenomena, such as the anisotropic magneto-Seebeck effect (AMSE) [5,12,22,26-29,37] and anomalous Nernst effect (ANE) [9,10,13,42], were discovered many decades ago, but its physics and materials science have not been developed sufficiently. This situation is being changed with rapid developments of spin caloritronics [24,32], where these phenomena are attracting renewed attention. In this stream, the observation of the anisotropic magneto-Peltier effect (AMPE) [37,41,45], the reciprocal of the AMSE, and the anomalous Ettingshausen effect (AEE) [39,41-43], the reciprocal of the ANE, in ferromagnetic metals has recently been reported.

The AMPE is a phenomenon that the Peltier coefficient depends on the angle $\theta_M$ between the directions of a charge current $\mathbf{J}_c$ and magnetization $\mathbf{M}$ in a ferromagnet. Due to the AMPE, the Peltier coefficient $\Pi$ in an isotropic ferromagnet exhibits the $\theta_M$ dependence satisfying the following equation:

$$\Pi(\theta_M) = \Pi_\perp + (\Pi_\parallel - \Pi_\perp) \cos^2\theta_M, \tag{1}$$

where $\Pi_{\parallel(\perp)}$ is the Peltier coefficient for the $\mathbf{M} \parallel \mathbf{J}_c$ ($\mathbf{M} \perp \mathbf{J}_c$) configuration. This symmetry is similar to that of the anisotropic magnetoresistance (AMR) in an isotropic ferromagnet [6]. If the finite AMPE appears, by forming a non-uniform magnetization configuration, the Peltier cooling and heating can be generated between the areas with different $\theta_M$ values even in the absence of junction structures and the magnitude of the resultant temperature change is proportional to $\Pi_\parallel - \Pi_\perp$. In 2018, the temperature change induced by the AMPE was directly observed in a polycrystalline Ni slab with U-shaped structure [41], where the Ni slab was uniformly magnetized and a charge current was applied along the U-shaped structure. This configuration allows us to realize the non-uniform $\theta_M$ distribution; the AMPE-induced temperature modulation appears around the corners of the U-shaped structure, i.e., the region between the area satisfying $\mathbf{M} \perp \mathbf{J}_c$ and the area satisfying $\mathbf{M} \parallel \mathbf{J}_c$ [Fig. 1(a)]. The AMPE has been observed not only in Ni but also in several ferromagnetic metals, such as $Ni_{95}Pt_5$ and $Ni_{95}Pd_5$ [41]. Importantly, the magnitude of the AMPE signal for $Ni_{95}Pt_5$ was found to be much greater than that for Ni, showing the important role of spin-orbit interaction in the AMPE [41]. Since the thermoelectric properties of the AMPE can be redesigned simply by changing the shape of ferromagnets or magnetization configurations, it provides new concepts in thermal management technologies for electronic and spintronic devices. Owing to the absence of junction structure, the AMPE-based devices can be easily integrated in nanoscale devices and integrated circuits where conventional thermoelectric devices cannot be integrated due to the complicated structure.

In a ferromagnet, the transverse thermoelectric effect called the AEE also appears. The AEE generates a heat current in the direction of the cross product of $\mathbf{J}_c$ and $\mathbf{M}$:

$$\mathbf{j}_{q,\text{AEE}} = \Pi_{\text{AEE}} \left( \mathbf{j}_c \times \mathbf{m} \right), \tag{2}$$

where $\mathbf{j}_{q,\text{AEE}}$, $\mathbf{j}_c$, $\mathbf{m}$, and $\Pi_{\text{AEE}}$ denote the heat-current density driven by the AEE, charge current density applied to a ferromagnet, unit vector of $\mathbf{M}$, and anomalous Ettingshausen coefficient, respectively [Fig. 1(a)] [39,41-43]. Therefore, when $\Pi_{\text{AEE}}$ is isotropic in a ferromagnet, the temperature gradient induced by the AEE should be proportional to $\sin\theta_M$. The AEE may also bring novel thermal management technologies because it enables magnetic control of a heat-current direction.

The understanding of the microscopic mechanism of the AMPE and AEE in ferromagnetic conductors and improvement of their thermoelectric conversion efficiency are crucial not only for the fundamental physics of spin caloritronics and thermoelectrics but also for the realization of nano-scale thermal energy engineering. However, despite their scientific and technological importance, few studies on the AMPE and AEE have been reported so far, and more comprehensive studies are necessary. Although the direct observation of the AMPE has been reported only for bulk ferromagnetic metal slabs, thin films make it easier to perform systematic measurements, including the composition dependence and the effect of alloy ordering [44,45], which accelerates research to find and develop materials exhibiting good thermoelectric conversion properties. Investigation of the AMPE in thin films is important also for integrating it in spintronic and spin-caloritronic devices for their thermal management. Significantly, in thin films formed on substrates, the magnitude of the AMPE-induced temperature modulation should depend not only on physical properties of the films but also on the film dimensions and thermal properties of the substrates due to heat loss from the films to the substrates [21,30,35]. The role of the substrates in the AEE experiments is also yet to be investigated systematically. The purpose of this study is to establish a procedure for the detection of the AMPE and AEE contributions in thin films and to clarify the dependences of the AMPE and AEE on the thickness of the films and thermal properties of the substrates.

In this paper, we report the observation of the AMPE and AEE in Ni thin films with changing the film thickness and substrate materials by means of the lock-in thermography (LIT) method [38-45,47-49]. We have demonstrated that the pure AMPE and AEE contributions, separated from other thermoelectric effects, can be extracted even in the thin films by measuring the magnetic-field-angle dependence of the LIT images. The field-angle dependence of the AMPE and AEE signals was confirmed to show the symmetries consistent with Eqs. (1) and (2), respectively. We also found that both the AMPE- and AEE-induced temperature modulations increase almost linearly with the increase in the Ni-film thickness and that the AMPE-induced temperature modulation decreases with the increase in thermal conductivity of the substrates, whereas the AEE-induced temperature modulation is nearly independent of the substrates' thermal conductivity. These detailed studies provide new insights to reveal the parameters that are relevant for optimizing the device geometry and material selection for practical applications of these phenomena and to construct their basic understanding.

This paper is organized as follows. In Sec. II, we explain the details of the experimental procedures and configurations for the measurements of the AMPE and AEE using the LIT method. In Sec. III, we report the observation of the AMPE- and AEE-induced temperature modulation in Ni films, followed by numerical calculations to explain the observed behaviors. The last Sec. IV is devoted to the conclusion of the present study.

## II. EXPERIMENTAL PROCEDURE

The samples used in this study are U-shaped Ni thin films of various thicknesses (50, 100, 150, and 200 nm) deposited on glass, gadolinium gallium garnet (GGG), and sapphire substrates with a size of $10 \times 10 \times 0.5$ mm$^3$ by using a RF magnetron sputtering system. Immediately before the deposition, the substrates were cleaned with acetone and ethanol using an ultrasonic cleaner. The RF power and Ar gas pressure for the Ni deposition were kept fixed at 50 W and 0.3 Pa, respectively. The deposition rate was 0.019 nm/s for all the samples. The Ni films were formed on the well-polished $10 \times 10$ mm$^2$ surface of the substrates and patterned into U-shaped structure by sputtering Ni through a metallic shadow mask, where the width of the area B$_{L/R/C}$ is 0.2 mm [Fig. 1(b)]. The thickness of the deposited films was measured using a stylus surface profilometer. To determine the saturation field of the films, AMR measurements were performed. It was found that the magnetization of the Ni films saturates under an in-plane magnetic field **H** with the magnitude of $H > 10$ Oe. In the U-shaped Ni film, when **H** is applied along the $x$ direction and **M** is aligned along the **H** direction, the AMPE signal appears around the corner C$_{L/R}$, the region between the area $B_{L/R}$ satisfying $\mathbf{M} \perp \mathbf{J}_c$ and the area $B_C$ satisfying $\mathbf{M} \parallel \mathbf{J}_c$, while the AEE signal appears in $B_{L/R}$ [Figs. 1(a) and 1(b)]. Therefore, the AMPE and AEE can be separated from each other by the position or spatial distribution of temperature-modulation signals. The magnitude of the AMPE and AEE signals increases with increasing $|H|$ following the magnetization curve of the Ni films and becomes constant after the saturation field. Importantly, the AMPE (AEE) signals exhibit the even (odd) dependence on the $H$ sign when $\mathbf{M}$ is along the $x$ direction [Eqs. (1) and (2)] [41]. In this paper, we focus on the bulk thermoelectric conversion properties and neglect interface/surface effects because the thickness of the Ni films used here is much larger than typical electrons' spin or magnon diffusion lengths.

The measurements of the AMPE and AEE in the Ni films were performed by using the LIT method [Fig. 1(c)] [38-45,47-49]. In the LIT experiments, we measure the spatial distribution of infrared radiation thermally emitted from the sample surface with high temperature and spatial resolutions, enabling quantitative measurements of the AMPE- and AEE-induced temperature modulation in the Ni films. To enhance infrared emissivity, the surface of the samples was coated with insulating black ink, of which the emissivity is > 0.95. During the LIT measurements, a square-wave-modulated AC charge current with the amplitude $J_c$, frequency $f$, and zero DC offset was applied to the Ni films and the sample was fixed on an Al block, which can be regarded as a heat bath. The thermal images oscillating with the same frequency as the input charge current were then extracted through Fourier analysis. The contribution of the thermoelectric effects ($\propto J_c$) can be separated from the Joule-heating background ($\propto J_c^2$) by extracting the first harmonic response of the thermal images, since the Joule heating generated in this condition is constant with time. As a result of the Fourier analysis, the obtained thermal images were transformed into the lock-in amplitude $A$ and phase $\phi$ images. The $A$ image shows the distribution of the magnitude of the current-induced temperature modulation and the $\phi$ image shows the distribution of the sign of the temperature modulation in addition to the time delay due to thermal diffusion, where $A$ and $\phi$ values are defined in the ranges of $A \geq 0$ and $0° \leq \phi < 360°$, respectively. The detected infrared radiation is converted into temperature information through the calibration method shown in Ref. [41]. The LIT images were measured while applying $\mathbf{H}$ with a magnitude of $H = 100$ Oe to the Ni films in the $x$-$y$ plane at an azimuthal angle $\theta_H$ from the $+x$ direction [Fig. 1(b)], where $|H|$ is much higher than the saturation field of the films. All the measurements were carried out at room temperature and atmospheric pressure.

The measurements of the magnetic-field-angle $\theta_H$ dependence of the temperature modulation enable the pure detection of the AMPE and AEE and the demonstration of their symmetries. To discuss the symmetries of the AMPE and AEE, it is necessary to note the relations between $\theta_M$ and $\theta_H$. As shown in Fig. 1(b), the relations are $\theta_M = 270° - \theta_H$, $90° - \theta_H$, and $180° - \theta_H$ in the areas $B_L$, $B_R$, and $B_C$ of the U-shaped structure, respectively. Solving the trigonometric relations in Eq. (1) [Eq. (2)], it was found that the temperature modulation induced by the AMPE (AEE) in the U-shaped ferromagnet is proportional to $\cos2\theta_H$ ($\cos\theta_H$). Hereafter, the symmetries of the AMPE and AEE are discussed in terms of $\theta_H$, not $\theta_M$.

## III. RESULTS AND DISCUSSION
### A. Fundamental measurements and analyses

In Figs. 2(a)-2(d), we show the $A$ and $\phi$ images for the U-shaped 100-nm-thick Ni film on the glass substrate for $\theta_H = 0°$, $90°$, $180°$, and $270°$, respectively, measured at $f = 25$ Hz and $J_c = 40$ mA. To determine the positions of the AMPE- and AEE-induced heat sources/sinks, the lock-in frequency $f$ was fixed at the maximum value, because the temperature broadening due to thermal diffusion is reduced by increasing $f$, where the maximum $f$ value for the frame size used in this study (640 × 512 pixels) is 25 Hz. A clear charge-current-induced temperature modulation was observed to appear in the Ni film at all the $\theta_H$ values. As discussed above, to separate the AMPE (AEE) contributions, we extracted the component showing the $\cos2\theta_H$ ($\cos\theta_H$) symmetry from the raw LIT images, where the LIT amplitude and phase of the $\cos2\theta_H$ ($\cos\theta_H$) component are denoted by $A_{\text{AMPE(AEE)}}$ and $\phi_{\text{AMPE(AEE)}}$, respectively. Figure 2(e) shows the $A_{\text{AMPE}}$ and $\phi_{\text{AMPE}}$ images for the U-shaped 100-nm-thick Ni film on the glass substrate at $\theta_H = 0°$. We observed clear temperature-modulation signals with the $\cos2\theta_H$ symmetry, which are generated predominantly around $C_{L/R}$ of the U-shaped structure and the sign of the temperature modulation at $C_L$ is opposite to that at $C_R$ because of the $\phi_{\text{AMPE}}$ difference of ~180°. This behavior is consistent with the characteristic of the AMPE in ferromagnetic metal slabs [41]. We also confirmed that the sign of the temperature modulation at $C_{L/R}$ for the Ni film is the same as that of the AMPE signals for the Ni slab [41]. The detailed behaviors of the AMPE in the Ni film will be discussed in Sec. IIIB with showing the $\theta_H$ dependence of the $A_{\text{AMPE}}$ and $\phi_{\text{AMPE}}$ images. In contrast, as shown in the $A_{\text{AEE}}$ and $\phi_{\text{AEE}}$ images in Fig. 2(f), the signal with the $\cos\theta_H$ symmetry is generated primarily on the area $B_{L/R}$ of the U-shaped structure, where $\mathbf{M} \perp \mathbf{J}_c$. The $\phi_{\text{AEE}}$ difference between $B_L$ and $B_R$ was observed to be ~180°. The behavior of this temperature modulation is consistent with the characteristic of the AEE in in-plane magnetized (IM) ferromagnetic metals [39,41-43]. The detailed behaviors of the AEE in the Ni film will be discussed in Sec. IIIC. Here, we also note that the background signal in the raw LIT images. In Figs. 2(a)-2(d), in addition to the AMPE and AEE signals, a complicated patchy pattern was observed to appear on the whole surface of the Ni film, although no patchy pattern signals were observed in the previous measurements using Ni slabs [41]. The appearance of the patchy pattern could be due to the polycrystalline nature of the thin films where the Peltier coefficient may have a different value depending on the quality and orientation of the grains. However, since such background coming from the conventional Peltier effect is independent of the magnetic field, it is eliminated in the AMPE and AEE images shown in Figs. 2(e) and 2(f) [see also the $H$-independent background images shown in Fig. 2(g)]. This procedure of extracting the pure AMPE and AEE signals can be extended to other systems where a complicated temperature modulation is generated due to the polycrystalline nature of the sample.

### B. Anisotropic magneto-Peltier effect

Here, we report the systematic measurements of the AMPE-induced temperature modulation in the Ni films. Figures 3(a) and 3(b) show the $\theta_H$ dependence of the $A_{\text{AMPE}}$ and $\phi_{\text{AMPE}}$ images, respectively, for the 100-nm-thick Ni film on the glass substrate at $f = 25$ Hz and $J_c = 40$ mA. We found that the AMPE signal at $C_{L/R}$ at $\theta_H = 90°$ is opposite in sign to that at $\theta_H = 0°$ and the signal disappears at $\theta_H = 45°$, where the disappearance of the AMPE-induced temperature modulation at $C_{L/R}$ is expected as both $B_{L/R}$ and $B_C$ have the same Peltier coefficient at $\theta_H = 45°$ [41]. Figure 3(e) shows $\Delta T_{\text{AMPE}}$ on $C_L$ and $C_R$ at $f = 25$ Hz and $J_c = 40$ mA as a function of $\theta_H$, where $\Delta T_{\text{AMPE}} = A_{\text{AMPE}}\cos\phi_{\text{AMPE}}$ represents the temperature modulation with sign information if the phase delay due to thermal diffusion is negligibly small. The $\Delta T_{\text{AMPE}}$ signal varies with $\theta_H$ in a $\cos2\theta_H$ pattern at both $C_L$ and $C_R$ and its sign reverses between $C_L$ and $C_R$ for each $\theta_H$ value. As discussed in Sec. II, this $\theta_H$ dependence of the temperature modulation is consistent with Eq. (1), confirming the symmetry of the AMPE. The behavior of the transverse response of the AMPE, i.e., a planar Ettingshausen effect, for the Ni films is also consistent with that for the Ni slab and previous expectations [41]; the magnitude of the temperature gradient along the $x$ ($y$) direction in $B_{L/R}$ ($B_C$) becomes minimum (maximum) at $\theta_H = 0°$ and $90°$ ($\theta_H = 45°$). We also found that the $f$ dependence of the AMPE-induced temperature modulation for the Ni films is similar to that for the Ni slabs without substrates [41]; the magnitude of $A_{AMPE}$ at $C_L$ and $C_R$ gradually increases with decreasing $f$ with accompanying the temperature broadening due to thermal diffusion [Figs. 3(c), 3(d), 3(f), and 3(g)]. All the experimental results shown in Fig. 3 conclude that the AMPE-induced temperature modulation is detectable even in thin film ferromagnets formed on substrates.

To understand the effect of thermal properties of the substrates on the AMPE-induced temperature modulation, we performed LIT measurements using Ni thin films on different substrates. Figures 4(a)-4(c) show the $A_{AMPE}$ and $\phi_{AMPE}$ images for the U-shaped 100-nm-thick Ni film on the glass, GGG, and sapphire substrates at $f$ = 25 Hz and $J_c$ = 40, 60, 75 mA, respectively. Although all the samples exhibit the clear AMPE-induced temperature modulation, its magnitude in the Ni films on the GGG and sapphire substrates was observed to be much smaller than that on the glass substrate. We found that the magnitude of the AMPE signals monotonically decreases with increasing the thermal conductivity of the substrates $\kappa_{sub}$ [see the $\kappa_{sub}$ dependence of $A_{AMPE}/j_c$ at $C_L$ of the Ni thin films in Fig. 4(h), where $j_c$ represents the charge current density applied to the Ni films, and the $\kappa_{sub}$ values are listed in Table I]. As shown in Figs. 4(d)-4(g), the $f$ dependence of the AMPE-induced temperature modulation for the thin films becomes weaker with increasing $\kappa_{sub}$. These behaviors are attributed to the heat loss from the Ni films to the substrates, which increases with increasing $\kappa_{sub}$ in this measurement condition, indicating that the magnitude of the AMPE-induced temperature modulation in thin films is determined by the ratio between the amount of heat sources/sinks generated by the AMPE and the heat loss to surroundings. In fact, we confirmed that the magnitude of the AMPE signals per unit charge current density increases almost linearly with increasing the thickness of the Ni films for all the substrate species [Fig. 4(i)]. These results are reproduced by the numerical calculations shown in Sec. IIID. Our findings conclude that the magnitude of the AMPE-induced temperature modulation in thin film devices is determined not only by thermoelectric properties of the films but also by the film dimensions and thermal boundary conditions and that substrates with lower thermal conductivity are better for obtaining larger AMPE signals in thin films. Importantly, the direct observation of the AMPE in thin films makes it possible to investigate the AMPE in various materials and conditions, because one can obtain the information on the sign and relative magnitude of the anisotropy of the Peltier coefficient even from thin film samples as long as the same substrate is used.

### C. Anomalous Ettingshausen effect

Now, we are in a position to investigate the AEE in thin films and clarify its behaviors. To confirm the symmetry of the AEE signal, we have systematically explored the $\theta_H$ dependence of the AEE-induced temperature modulation. Figures 5(a) and 5(b) show the $A_{AEE}$ and $\phi_{AEE}$ images, respectively, for various values of $\theta_H$ for the 100-nm-thick Ni film on the sapphire substrate at $f$ = 25 Hz and $J_c$ = 75 mA. We found that the AEE signal at $B_{L/R}$ decreases with increasing $\theta_H$ and disappears when $\theta_H = 90°$. The AEE signal at $B_C$ was observed to follow opposite trend to that at $B_{L/R}$; it becomes minimum (maximum) at $\theta_H = 0°$ ($\theta_H = 90°$). Figure 5(e) shows $\Delta T_{AEE}$ on $B_L$ and $B_R$ at $f$ = 25 Hz and $J_c$ = 75 mA as a function of $\theta_H$, where $\Delta T_{AEE} = A_{AEE}\cos\phi_{AEE}$. The $\Delta T_{AEE}$ signal varies with $\theta_H$ in a $\cos\theta_H$ pattern [Fig. 5(e)] at both $B_L$ and $B_R$ and its sign reverses between $B_L$ and $B_R$ for each $\theta_H$ value. As discussed in Sec. II, this $\theta_H$ dependence of the temperature modulation is consistent with Eq. (2) (recall the relations $\theta_M = 270° - \theta_H$, $90° - \theta_H$, and $180° - \theta_H$ for $B_L$, $B_R$, and $B_C$, respectively). Here, we note that, in the $A_{AEE}$ and $\phi_{AEE}$ images, the contribution coming from the ordinary Ettingshausen effect is negligibly small because of the small magnitude of the external magnetic field [41]. The $f$ dependence of the $A_{AEE}$ and $\phi_{AEE}$ images is also consistent with the recently-observed behavior of the AEE in IM ferromagnetic metal films [39,41-43]; unlike AMPE, the AEE signal at $B_{L/R}$ remains almost constant with $f$ and does not show any temperature broadening from the areas [Figs. 5(c), 5(d), 5(f), and 5(g)]. This result shows that the AEE-induced temperature modulation immediately reaches a steady state, a behavior which is similar to that of the spin Peltier effect where the spin-current-induced temperature modulation immediately reaches a steady state due to the presence of dipolar heat sources [38]. The difference in the $f$ dependence between the AMPE and AEE arises from the aspect ratio of the samples. The AEE induces a heat current along the thickness direction of the Ni films, while the AMPE along the lateral directions. Since the thickness of the Ni films is three orders of magnitude smaller than the lateral size, the AEE-induced temperature change reaches the steady state much faster than the AMPE-induced one, resulting in the constant $f$ dependence of the AEE signals.

Figures 6(a)-6(c) show the $A_{AEE}$ and $\phi_{AEE}$ images for the U-shaped 100-nm-thick Ni films on the glass, GGG, and sapphire substrates at $f = 25$ Hz and $J_c = 40, 60, 75$ mA, respectively. We observed the AEE signals with the same sign and similar magnitude in all the samples. As shown in Figs. 6(d)-6(g), for all the substrates, the AEE-induced temperature modulation is almost independent of $f$. Significantly, our results clearly show that the $\kappa_{sub}$ dependence of $A_{AEE}/j_c$ for the Ni films is much weaker than that of the AMPE signals [compare Fig. 6(h) with Fig. 4(h)], indicating that the AEE-induced temperature modulation in the IM configuration is not affected by the thermal properties of the substrates [50]. This is unlike the AMPE-induced temperature modulation where the AMPE signal decreases with the increase in $\kappa_{sub}$, confirming the different symmetry of the AMPE and AEE. As further discussed in Sec. IIID, the observed behaviors can be reproduced by taking the difference in heat-source/sink distributions induced by these phenomena into account. It should also be noted that, in Refs. [39,42], the AEE-induced temperature modulation in the perpendicularly magnetized (PM) configuration, where the heat current is generated along the in-plane direction, was shown to be strongly affected by the heat loss to the substrate, which is different from the behavior for the AEE in the IM configuration. Therefore, the AEE signals in the PM configuration is expected to be increased with decreasing $\kappa_{sub}$ in a similar manner to the $\kappa_{sub}$ dependence of the AMPE signals. In fact, although the Ni films on the glass substrate with low $\kappa_{sub}$ exhibit the clear IM-AEE features, the signals are contaminated by the PM-AEE contribution due to the small but finite perpendicular component of **M**, or non-uniformity of **H** [Fig. 6(a)]. In contrast, the pure IM-AEE signals were observed in the Ni films on the GGG and sapphire substrates having relatively-high $\kappa_{sub}$ [Figs. 6(b) and 6(c)], where the parasitic PM-AEE is suppressed.

Figure 6(i) shows that the AEE signals also increase linearly with increasing the film thickness for all the substrates. This thickness dependence of the AEE signals is consistent with the results in Ref. [43] and explained simply by the facts that the out-of-plane heat current induced by the IM-AEE is constant in the Ni films and that the resultant temperature difference is proportional to the integral of the heat current over the Ni thickness.

### D. Numerical calculations

To confirm the observed dependences of the AMPE and AEE signals in the Ni films on the film thickness and $\kappa_{sub}$, we calculated the AMPE- and AEE-induced temperature distributions by using a finite element method [51]. The temperature modulation due to the AMPE and AEE are evaluated in the corresponding two-dimensional cross sections (*x-z* plane) of the U-shaped strip by disregarding the *y* direction [Figs. 7(a) and 7(b)]. The model used for the calculations of the AMPE-induced (AEE-induced) temperature modulation consists of a Ni film with a width of 1.0 mm (two separated Ni films with a width of 0.2 mm) and a thickness of 50, 100, 150, or 200 nm, a substrate with a width of 10 mm and a thickness of 0.5 mm, and a black-ink layer with a width of 10 mm and a thickness of 10 μm, which corresponds to the cross-section of the sample used for the experiments across $C_{L/R}$ ($B_{L/R}$) [compare Figs. 7(a) and 7(b) with Fig. 1(b) and note that the two separated Ni films in the model used for the AEE calculations represent the two leg parts of the U-shaped Ni film]. Since the LIT experiments were performed in the frequency domain, we numerically solved the heat equation in the form of

$$Ae^{-i\phi} = \frac{1}{2\pi i f C\rho}\left[\nabla \cdot \left(\kappa \nabla A e^{-i\phi}\right) + Q(r)\right], \tag{3}$$

where $C$, $\rho$, $\kappa$, and $Q(r)$ denote the specific heat, density, thermal conductivity, and heat sources. As schematically depicted in Figs. 7(a) and 7(b), $Q(r)$ is approximately set based on the previous calculations [39,41]. Here, the AMPE generates a single heat source $+Q$ or heat sink $-Q$ around the corners of the U-shaped Ni film, where the lateral heat current is discontinuous due to the anisotropy of the Peltier coefficient [Fig. 7(a)]. In contrast, the AEE generates a heat current along the *z* direction; this situation is represented by the model in which the symmetric pairs of $+Q$ and $-Q$ (dipolar heat sources) are put at the top/bottom surfaces of the Ni films because the vertical heat current is discontinuous at the top/bottom surfaces [Fig. 7(b)]. The boundary conditions are $(-\kappa \nabla A) \cdot \mathbf{n} = \alpha_{air} A$ for both the model boundaries, where **n** is the surface normal vector directing the outside and $\alpha_{air}$ = 10 Wm$^{-2}$K$^{-1}$ [41] is the heat loss factor to the air, except the bottom of the substrate connected to the heat bath at which $A = 0$ is assumed. The material parameters used for the calculations are listed in Table I. For the calculations, we assumed $C\rho$ = 1.0 ×10$^6$ Jm$^{-3}$K$^{-1}$ and $\kappa$ = 0.5 Wm$^{-1}$K$^{-1}$ for the black-ink layer (note that, in Ref. [39], the similar numerical calculation using these values well reproduces the experimental results of the AEE). We note that, in the numerical calculations, the interfacial thermal resistance between the Ni films and substrates is neglected because its contribution is much smaller than the thermal resistance of the submillimeter-thick substrate [52-54].

To understand the behaviors of the AMPE in thin films on substrates, we calculated the temperature distribution in the model shown in Fig. 7(a), where the width of the heat source and sink is set to be 0.2 mm. Calculated $\Delta T_{AMPE}$ (= $A_{AMPE}\cos\phi_{AMPE}$) images show remarkable broadening of the AMPE-induced temperature modulation in the substrates. Importantly, the resultant temperature distribution is strongly dependent on thermal properties of the substrate; as shown in Fig. 7(c), the magnitude of the calculated AMPE-induced temperature modulation for the sapphire substrate is much smaller than that for the glass substrate. In Fig. 7(e), we plot the calculated values of $A_{AMPE}$ on the top surface of the black ink around the center of the positive heat source as a function of $\kappa_{sub}$, showing that the magnitude of $A_{AMPE}$ decreases with increasing $\kappa_{sub}$ and the behavior is consistent with the experimental results. The relationship between the Ni-film thickness and the AMPE-induced temperature modulation is also reproduced by the numerical calculations for all the substrates [Fig. 7(g)].

Next, we show the calculated temperature distribution induced by the AEE. In the model shown in Fig. 7(b), the thickness of the heat sources and sinks was set to be 1% of the film thickness and the width is the same as that of the Ni films (= 0.2 mm). The polarity of the dipolar heat sources is reversed between the left and right Ni films according to the symmetry of the AEE in our experiments [compare Fig. 7(b) with Fig. 1(a)]. The calculated temperature distributions show that the AEE-induced temperature modulation appears only in the vicinity of the Ni films and that the magnitude of the temperature modulation on the top surface of the sample is much greater than that inside the substrate, a situation which is similar to the temperature distribution induced by the spin Peltier effect [38]. Owing to this temperature distribution, the temperature modulation induced by the AEE in the IM configuration is hardly affected by the thermal properties of the substrates. In fact, as shown in Figs. 7(d) and 7(f), the magnitude of the calculated $A_{AEE}$ values on the surface of the black ink is almost independent of the substrate species, consistent with the experimental results (note that the small decrease in the AEE-induced temperature modulation for the glass sample is due to the situation that the heat loss to the substrate through the lateral edges of the Ni layer is not negligibly small as a result of comparable thermal conductivity between glass and the black-ink layer [55]). The relationship between the Ni-film thickness and the AEE-induced temperature modulation, demonstrated experimentally, is also reproduced by the numerical calculations for all the substrates [Fig. 7(h)].

## IV. SUMMARY

We investigated the AMPE and AEE in polycrystalline Ni thin films of varied thickness (50, 100, 150, and 200 nm) deposited on glass, GGG, and sapphire substrates by using the LIT technique. We demonstrated a procedure to extract the pure AMPE- and AEE-induced temperature-modulation contributions by separating them from each other and from the background due to the conventional Peltier effect. Our results show that the Ni films exhibit the clear current-induced temperature modulation showing the $\cos2\theta_H$ ($\cos\theta_H$) symmetry, confirming the presence of the AMPE (AEE) signals. It was found that both the AMPE- and AEE-induced temperature modulation increase linearly with increase in the film thickness and that the AMPE-induced temperature modulation increases with decreasing the frequency of the charge current applied to the Ni films, whereas the AEE-induced temperature modulation is almost independent of the frequency. The AMPE-induced temperature modulation was observed to decrease with increasing thermal conductivity of the substrates due to the heat loss from the films to the substrates, whereas the AEE-induced temperature modulation is almost independent of the thermal conductivity, which are reproduced by the numerical calculations based on a two-dimensional finite element method. These results establish the importance of the sample dimensions and substrate thermal property in determining the behaviors of the AMPE and AEE in thin films, highlighting the different symmetry of the two effects. Importantly, the behaviors of the AMPE and AEE revealed by this study are not limited to the LIT experiments; the substrate and thickness dependences of the temperature modulation induced by these phenomena have to be taken into account irrespective of measurement techniques.

## ACKNOWLEDGEMENT

The authors thank A. Miura, S. Daimon, S. Ota, D. Chiba, Y. Sakuraba, and T. Seki for valuable discussions and T. Yamazaki for technical supports. This work was supported by CREST "Creation of Innovative Core Technologies for Nano-enabled Thermal Management" (JPMJCR17I1) from JST, Japan; Grant-in-Aid for Scientific Research (S) (JP18H05246) and Grant-in-Aid for Early-Career Scientists (JP18K14116) from JSPS KAKENHI, Japan; and the NEC Corporation.

of the Ni layer is not negligibly small. This situation happens when $\kappa_{\text{sub}}$ is low. When the contribution of the heat flow from the lateral edges to the substrate bottom through the black-ink layer increases, the temperature at the Ni/substrate interface is no longer maintained. As a result, the surface temperature modulation decreases and opposite temperature modulation appears at the bottom of the Ni layer, resulting in the small temperature change near the substrate in Fig. 7(d). We also note that this substrate dependence decreases when the width of the Ni layer increases and disappears when the black-ink layer (or thermal conductor surrounding the Ni layer) is absent.

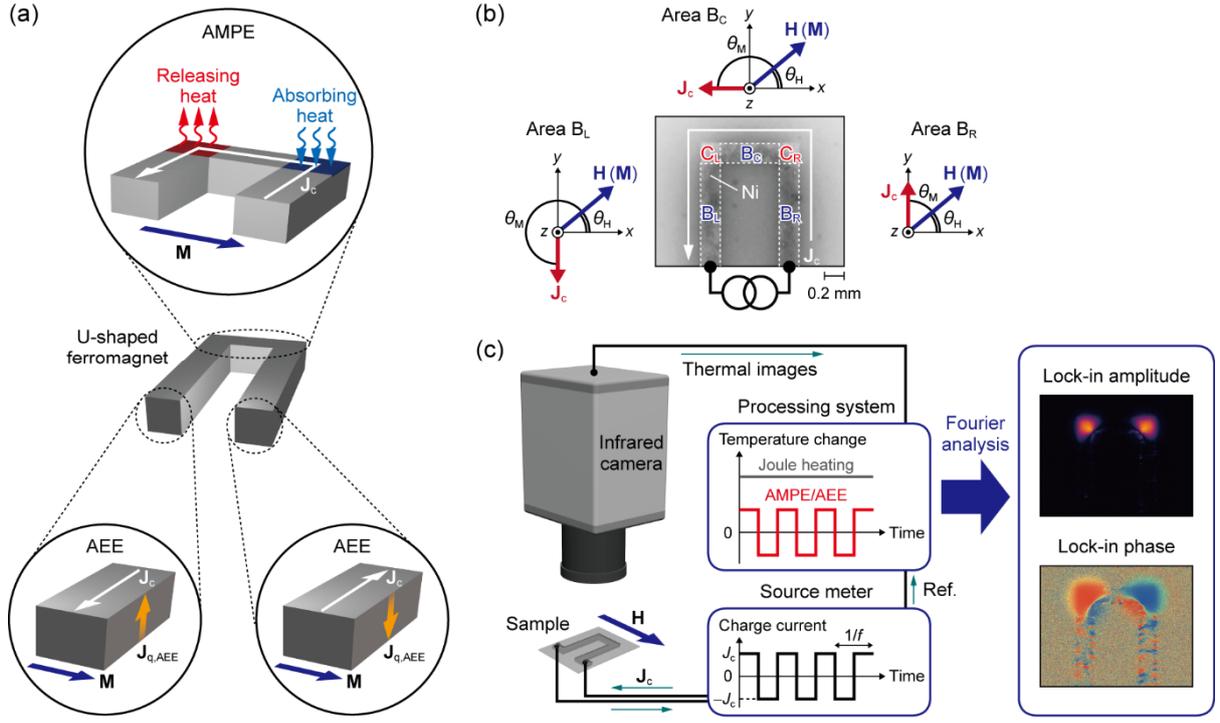

FIG. 1. (a) Schematic illustration of the U-shaped ferromagnetic metal used for measuring the AMPE and AEE. **M**, **J**$_c$, and **J**$_{q,\text{AEE}}$ denote the magnetization vector, charge current applied to the ferromagnetic metal (with the magnitude $J_c$), and heat current induced by the AEE. (b) Steady-state infrared image of the U-shaped Ni thin film with black-ink coating and the definitions of $\theta_M$ and $\theta_H$. $\theta_H$ ($\theta_M$) represents the angle between the magnetic field vector **H** (with the magnitude $H$) and the $+x$ direction (between **J**$_c$ and **M**), where the relations between $\theta_M$ and $\theta_H$ are defined when **M** is aligned along **H**. The area B$_{L/C/R}$, surrounded by white dotted lines, and the area C$_{L/R}$, the corner of the U-shaped structure, are the labels representing the parts of the Ni film. (c) Lock-in thermography (LIT) method used for the measurements of the AMPE and AEE. $f$ denotes the frequency of the charge current, i.e., the lock-in frequency. All the LIT measurements were performed at $|H|$ = 100 Oe.

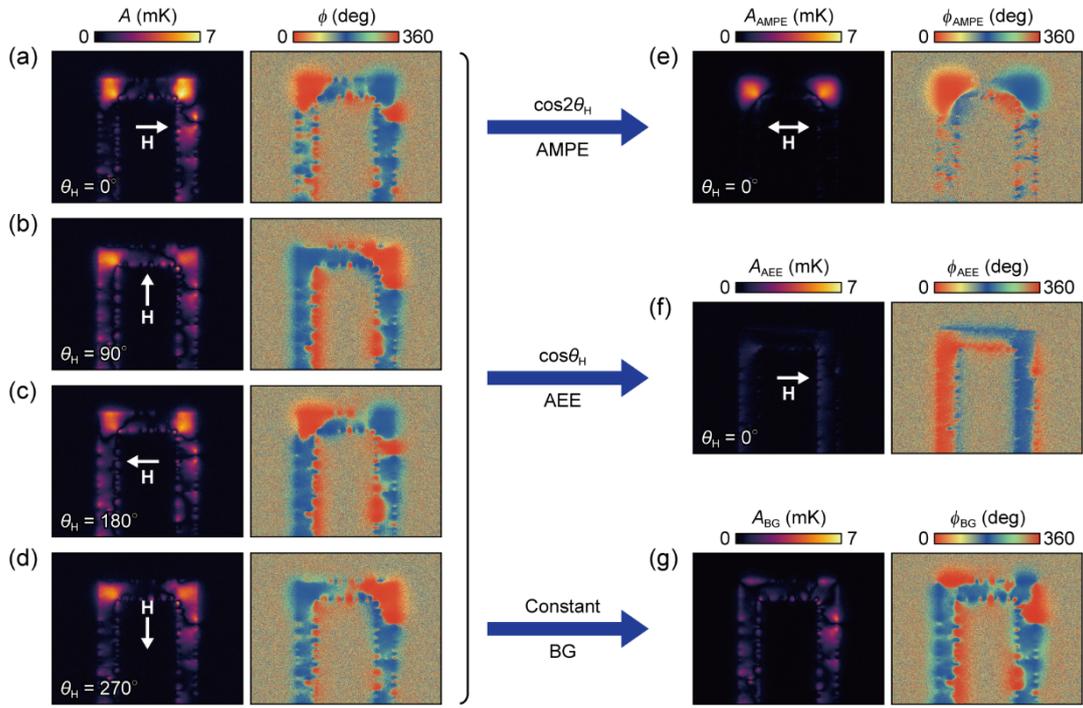

FIG. 2. (a)-(d) Lock-in amplitude $A$ and phase $\phi$ images for the U-shaped 100-nm-thick Ni film on the glass substrate at $\theta_H = 0°$, $90°$, $180°$, and $270°$, measured at $J_c = 40$ mA and $f = 25$ Hz. (e) $A_{AMPE}$ and $\phi_{AMPE}$ images for the U-shaped 100-nm-thick Ni film on the glass substrate at $\theta_H = 0°$, $J_c = 40$ mA, and $f = 25$ Hz. (f) $A_{AEE}$ and $\phi_{AEE}$ images for the U-shaped 100-nm-thick Ni film on the glass substrate at $\theta_H = 0°$, $J_c = 40$ mA, and $f = 25$ Hz. The LIT amplitude and phase with the $\cos2\theta_H$ ($\cos\theta_H$) symmetry are denoted by $A_{AMPE(AEE)}$ and $\phi_{AMPE(AEE)}$, respectively. (g) $A_{BG}$ and $\phi_{BG}$ images for the U-shaped 100-nm-thick Ni film on the glass substrate at $J_c = 40$ mA and $f = 25$ Hz. $A_{BG}$ ($\phi_{BG}$) denotes the amplitude (phase) of the temperature modulation due to the $H$-independent background.

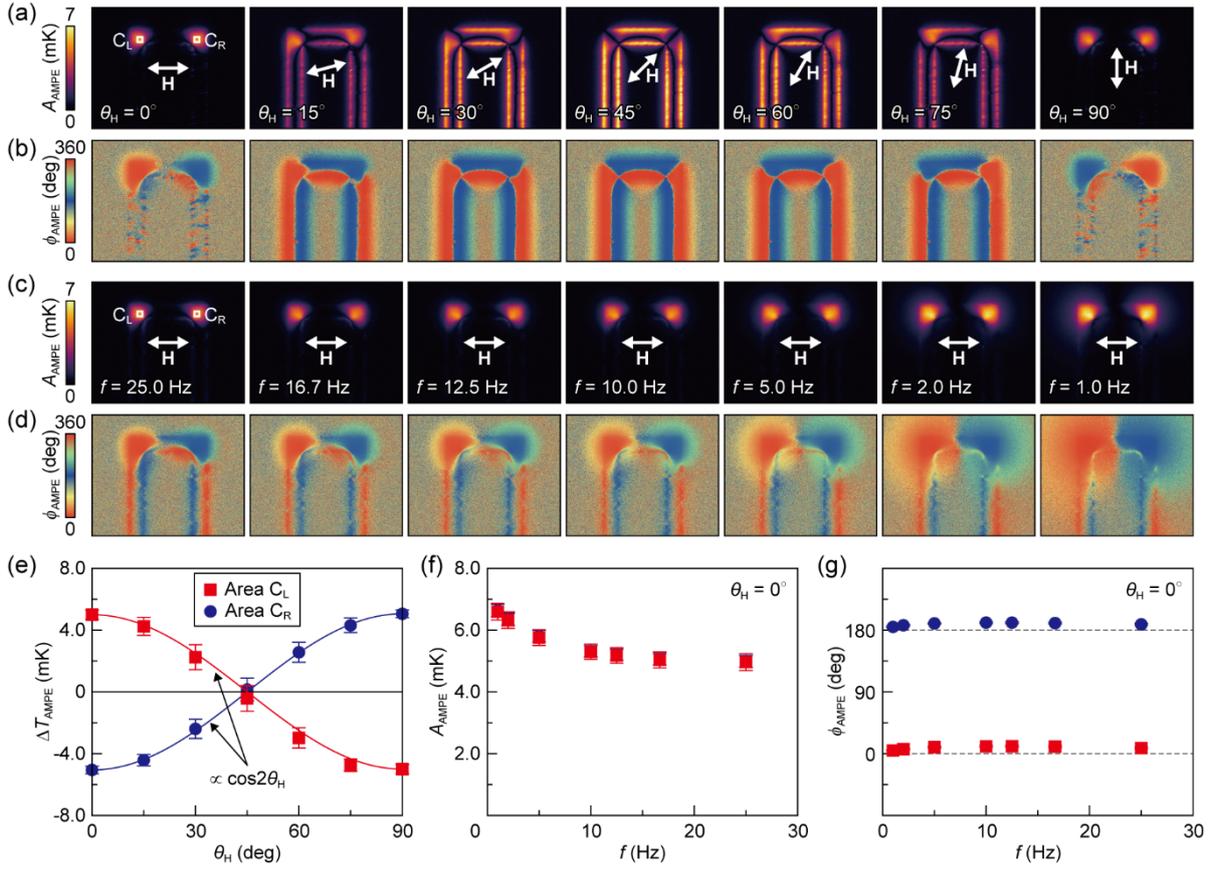

FIG. 3. (a),(b) $A_{\text{AMPE}}$ and $\phi_{\text{AMPE}}$ images for the U-shaped 100-nm-thick Ni film on the glass substrate for various values of $\theta_{\text{H}}$, measured at $J_\text{c} = 40$ mA and $f = 25$ Hz. (c),(d) $A_{\text{AMPE}}$ and $\phi_{\text{AMPE}}$ images for the U-shaped 100-nm-thick Ni film on the glass substrate for various values of $f$, measured at $\theta_{\text{H}} = 0°$ and $J_\text{c} = 40$ mA. (e) $\theta_{\text{H}}$ dependence of $\Delta T_{\text{AMPE}}$ on the area $C_{\text{L/R}}$ for the U-shaped 100-nm-thick Ni film on the glass substrate at $J_\text{c} = 40$ mA and $f = 25$ Hz. The solid curves show the $\cos 2\theta_{\text{H}}$ functions for the comparison with the experimental results. (f),(g) $f$ dependence of $A_{\text{AMPE}}$ and $\phi_{\text{AMPE}}$ on $C_{\text{L/R}}$ for the U-shaped 100-nm-thick Ni film on the glass substrate, measured at $\theta_{\text{H}} = 0°$ and $J_\text{c} = 40$ mA. The data points in (e)-(g) are obtained by averaging the $A_{\text{AMPE}}$ and $\phi_{\text{AMPE}}$ values on $C_{\text{L/R}}$ defined by the squares in (a) and (c). The error bars represent the standard deviation of the data in the corresponding squares.

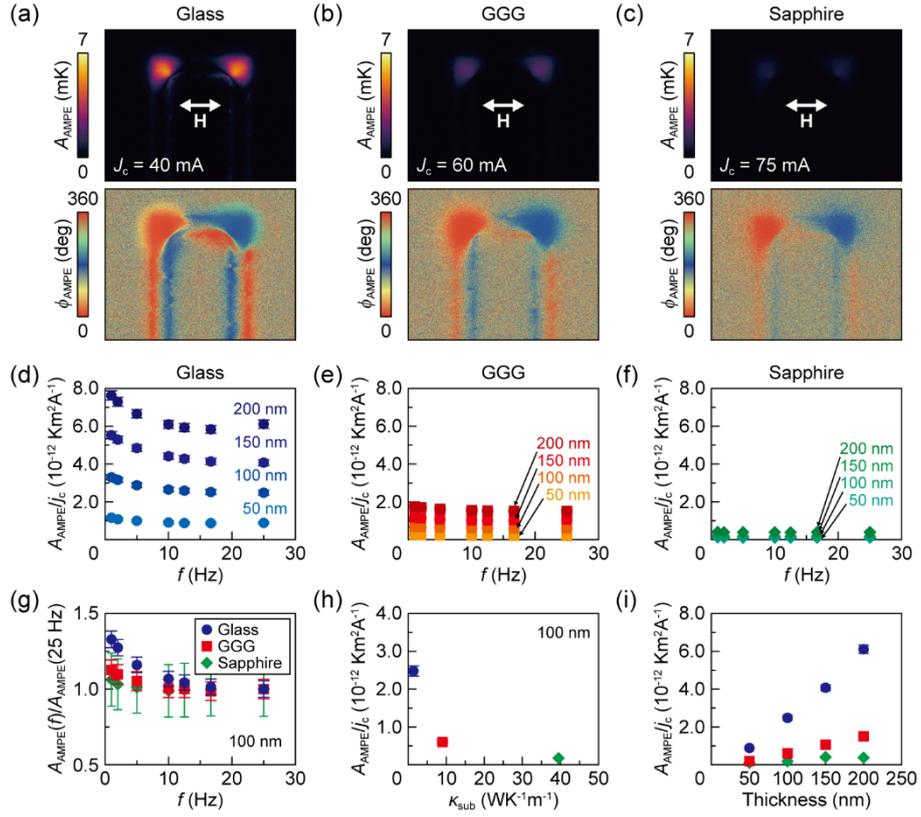

FIG. 4. (a)-(c) $A_{\text{AMPE}}$ and $\phi_{\text{AMPE}}$ images for the U-shaped 100-nm-thick Ni film on the glass, GGG and sapphire substrates, measured at $f = 25$ Hz and $J_c = 40, 60, 75$ mA, respectively. (d)-(f) $f$ dependence of $A_{\text{AMPE}}/j_c$ on $C_L$ for the Ni films with various thicknesses on the glass, GGG and sapphire substrates. $j_c$ represents the charge current density applied to the Ni films. (g) $f$ dependence of $A_{\text{AMPE}}(f)/A_{\text{AMPE}}(25\text{ Hz})$ for the U-shaped 100-nm-thick Ni film on the glass, GGG, and sapphire substrates. (h) $\kappa_{\text{sub}}$ dependence of $A_{\text{AMPE}}/j_c$ on $C_L$ for the 100-nm-thick Ni films on the glass, GGG, and sapphire substrates at $f = 25$ Hz. $\kappa_{\text{sub}}$ represents the thermal conductivity of the substrates. (i) Ni-film thickness dependence of $A_{\text{AMPE}}/j_c$ on $C_L$ for the Ni films on the glass, GGG, and sapphire substrates at $f = 25$ Hz. The error bars represent the standard deviation of the data in the corresponding squares. All the data shown in this figure were measured at $\theta_H = 0°$.

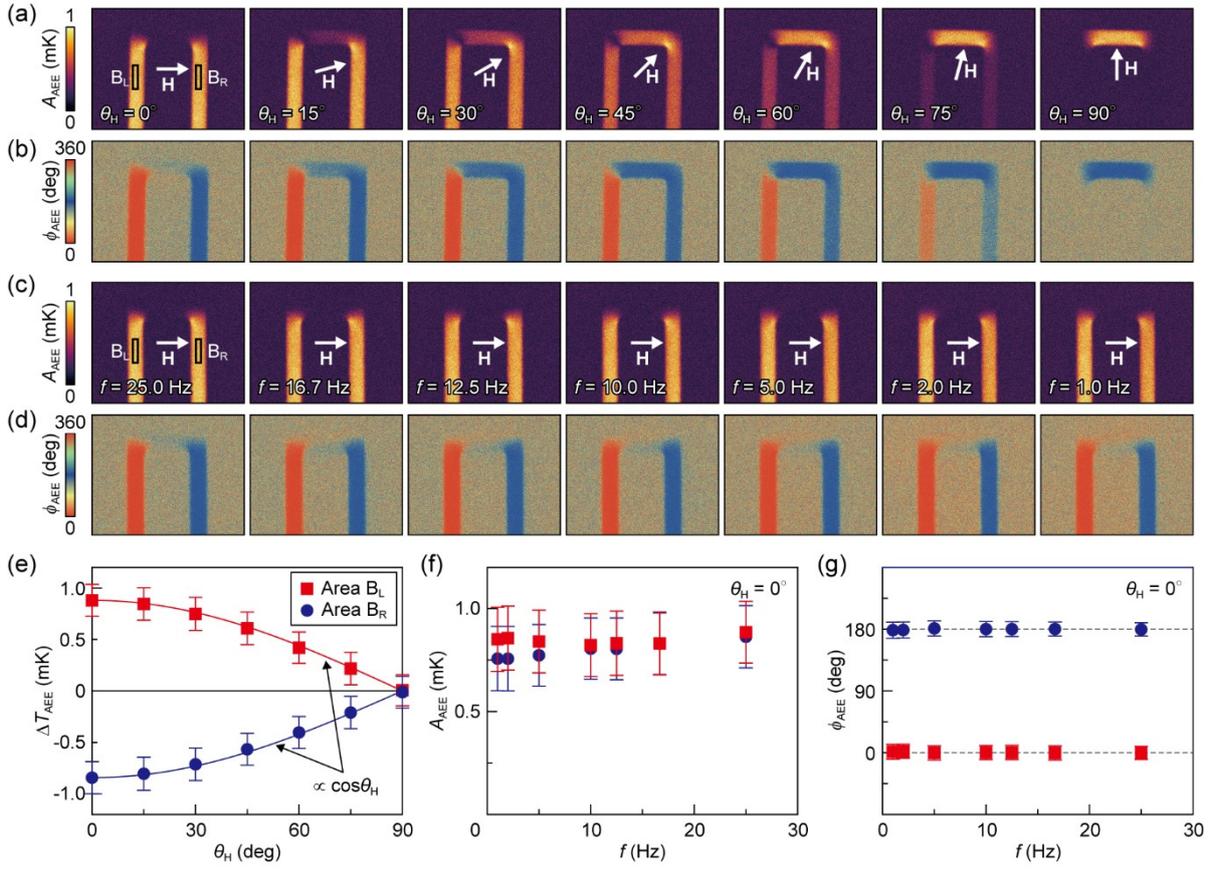

FIG. 5. (a),(b) $A_{AEE}$ and $\phi_{AEE}$ images for the U-shaped 100-nm-thick Ni film on the sapphire substrate for various values of $\theta_H$, measured at $J_c = 75$ mA and $f = 25$ Hz. (c),(d) $A_{AEE}$ and $\phi_{AEE}$ images for the U-shaped 100-nm-thick Ni film on the sapphire substrate for various values of $f$, measured at $\theta_H = 0°$ and $J_c = 75$ mA. (e) $\theta_H$ dependence of $\Delta T_{AEE}$ on the area $B_{L/R}$ for the U-shaped 100-nm-thick Ni film on the sapphire substrate at $J_c = 75$ mA and $f = 25$ Hz. The solid curves show the $\cos\theta_H$ functions for the comparison with the experimental results. (f),(g) $f$ dependence of $A_{AEE}$ and $\phi_{AEE}$ on $B_{L/R}$ for the U-shaped 100-nm-thick Ni film on the sapphire substrate, measured at $\theta_H = 0°$ and $J_c = 75$ mA. The data points in (e)-(g) are obtained by averaging the $A_{AEE}$ and $\phi_{AEE}$ values on $B_{L/R}$ defined by the squares in (a) and (c). The error bars represent the standard deviation of the data in the corresponding squares.

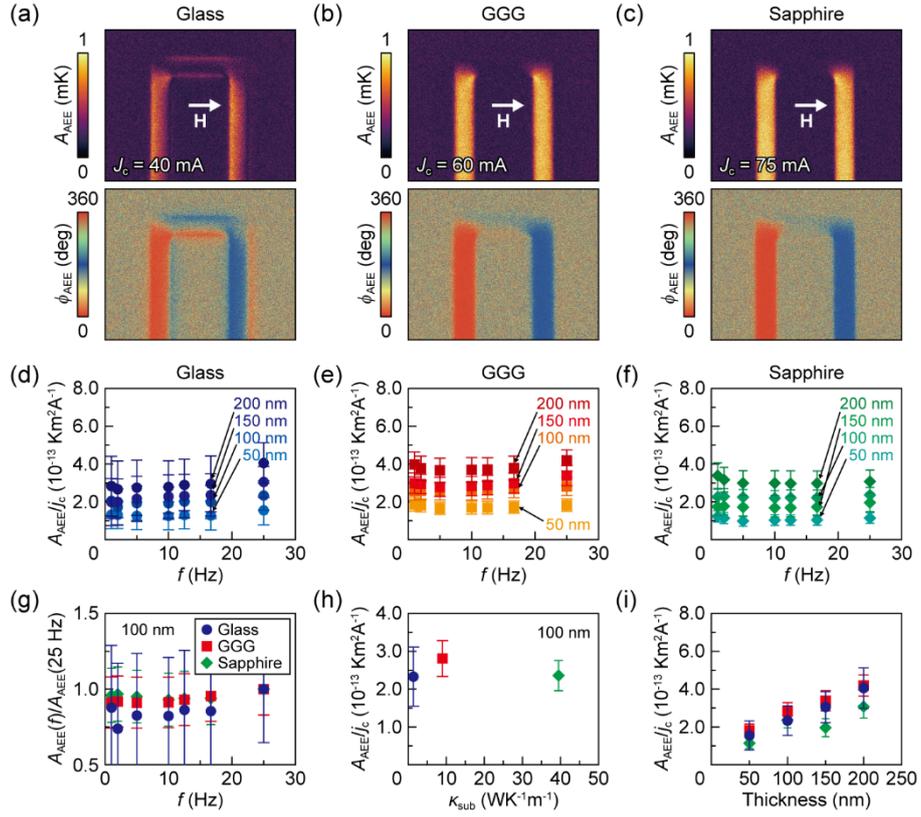

FIG. 6. (a)-(c) $A_{AEE}$ and $\phi_{AEE}$ images for the U-shaped 100-nm-thick Ni film on the glass, GGG and sapphire substrates, measured at $f = 25$ Hz and $J_c = 40, 60, 75$ mA, respectively. (d)-(f) $f$ dependence of $A_{AEE}/j_c$ on $B_L$ for the Ni films with various thicknesses on the glass, GGG and sapphire substrates. (g) $f$ dependence of $A_{AEE}(f)/A_{AEE}(25$ Hz) for the U-shaped 100-nm-thick Ni film on the glass, GGG, and sapphire substrates. (h) $\kappa_{sub}$ dependence of $A_{AEE}/j_c$ on $B_L$ for the 100-nm-thick Ni films on the glass, GGG, and sapphire substrates at $f = 25$ Hz. (i) Ni-film thickness dependence of $A_{AEE}/j_c$ on $B_L$ for the Ni films on the glass, GGG, and sapphire substrates at $f = 25$ Hz. The error bars represent the standard deviation of the data in the corresponding squares. All the data shown in this figure were measured at $\theta_H = 0°$.

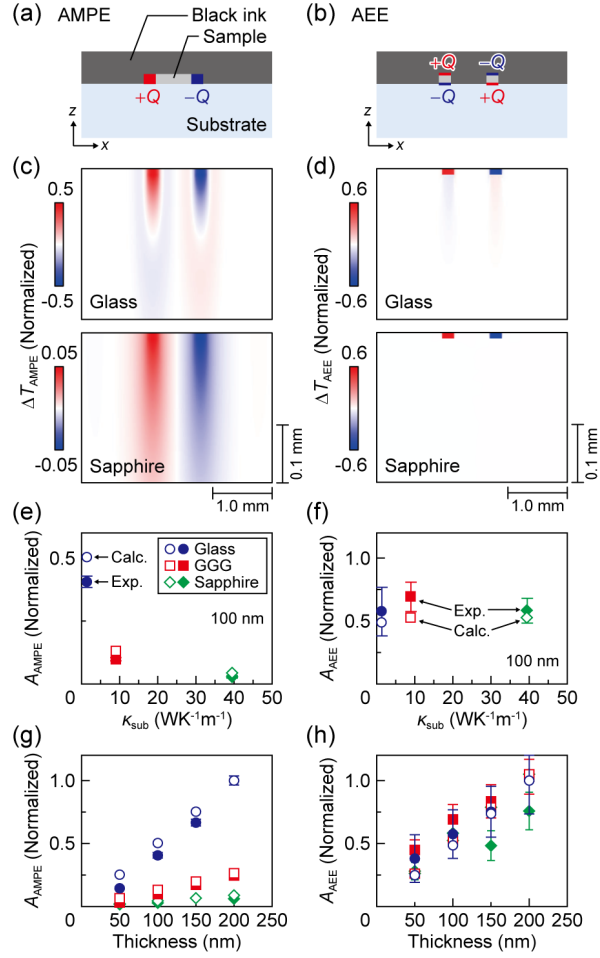

FIG. 7. (a),(b) Schematic illustrations of the model structure and the location of heat sources for the AMPE and AEE, used for the finite element calculations. (c) Calculated $\Delta T_{AMPE}$ (= $A_{AMPE}\cos\phi_{AMPE}$) distributions in the $z$-$x$ plane of the 100-nm-thick Ni film on the glass and sapphire substrates at $f$ = 25 Hz. (d) Calculated $\Delta T_{AEE}$ (= $A_{AEE}\cos\phi_{AEE}$) distributions in the $z$-$x$ plane of the 100-nm-thick Ni film on the glass and sapphire substrates at $f$ = 25 Hz. The images in (c) and (d) show the $\Delta T_{AMPE}$ and $\Delta T_{AEE}$ distributions near the top center of the black-ink/Ni-film/substrate models, where the heat sources are located near the top of the images. Note that the $\Delta T_{AMPE}$ and $\Delta T_{AEE}$ images are 10 times magnified in the $z$ direction. (e),(f) $\kappa_{sub}$ dependence of the calculated and experimental values of $A_{AMPE}$ and $A_{AEE}$ for the 100-nm-thick Ni films on the glass, GGG, and sapphire substrates at $f$ = 25 Hz. (g),(h) Ni-film thickness dependence of the calculated and experimental values of $A_{AMPE}$ and $A_{AEE}$ for the Ni films on the glass, GGG, and sapphire substrates at $f$ = 25 Hz. The calculated and experimental data in (e)-(h) were normalized by the $A_{AMPE/AEE}$ values for the 200-nm-thick Ni film on the glass substrate.

TABLE I. Materials parameters for numerical calculations.

|  | Ni | Glass | GGG | Sapphire |
|---|---|---|---|---|
| Thermal conductivity (Wm$^{-1}$K$^{-1}$) [56-58] | 26 | 1.4 | 9.0 | 39.5 |
| Density (g cm$^{-3}$) | 8.9 | 2.2 | 7.1 | 4.0 |
| Specific heat (Jg$^{-1}$K$^{-1}$) | 0.44 | 0.76 | 0.38 | 0.76 |